\documentclass[twocolumn,prd,aps,superscriptaddress,tightenlines,floatfix]{revtex4}
\usepackage{graphicx}
\usepackage{bm}

\newcommand{\xslash}[1]{{\rlap{$#1$}/}}

\newcommand{\me}[3]{\ensuremath{\left\langle{#1}\vphantom{#2 #3}
\right|{#2}\left|\vphantom{#1 #2}{#3}\right\rangle}}

\newcommand{\rme}[3]{\ensuremath{\left\langle\!\langle{#1}\vphantom{#2 #3}
\right|{#2}\left|\vphantom{#1 #2}{#3}\right\rangle\!\rangle}}

\newcommand{\nn}{\nonumber \\ }

\def\abs#1{ \left| #1 \right| }
\def\ket#1{ \left| #1 \right\rangle }
\def\bra#1{ \left\langle #1 \right| }

\begin{document}

\title{The Orthopositronium Decay Spectrum using NRQED}

\author{Aneesh V.~Manohar}
\affiliation{Department of Physics, University of California at San Diego,
  La Jolla, CA 92093\vspace{4pt} }

\author{Pedro Ruiz-Femen\'\i a}
\affiliation{Instituto de F\'\i sica Corpuscular, Universitat de Val\`encia -
CSIC, Apdo. 22085, 46071 Valencia, Spain\vspace{4pt} }
\date{November 2003}

\begin{abstract}
The  photon spectrum in orthopositronium $\to 3 \gamma$ decay is computed using effective field theory methods. For energetic photons, the spectrum agrees with the Ore-Powell result, but deviates from it when the photon energy is comparable to the positronium binding energy. The decay spectrum in this region depends on a positronium structure function, which is computed in this paper. At still lower energies the photon spectrum is dominated by the parapositronium resonance contribution. Our results are compatible with Low's theorem on soft photon emission.
\end{abstract}

\maketitle

\section{Introduction}

Since its discovery in 1951~\cite{deutsch} Positronium has become a  high precision laboratory for testing QED. A detailed study of the properties of the $e^+e^-$ spin triplet (orthopositronium: o-Ps) and spin singlet (parapositronium: p-Ps) states has provided impressive confirmation of QED radiative corrections.

Recently, it was pointed out~\cite{pestieau} that the calculation of the o-Ps decay spectrum, first performed by Ore and Powell~\cite{OrePowell}, is in apparent contradiction with Low's theorem, i.e.\ that the soft photon spectrum is not consistent with QED gauge invariance in combination with analyticity. While Low's theorem applied to this decay requires that the spectrum vanishes as $E_\gamma^3$, with $E_\gamma$ the energy of the radiated photon, the standard Ore-Powell calculation predicts a ${\cal O}(E_\gamma)$ behavior for $E_\gamma \to 0$. According to the authors of Ref.~\cite{pestieau}, this failure is traced back to the different radiation properties of the free charged leptons which are used as the asymptotic states in typical o-Ps$\to 3\gamma$ calculations, and those of positronium which is a neutral spin-1 boson.

In this paper we compute the orthopositronium $\to 3 \gamma$ decay spectrum using
non-relativistic effective field theory (NREFT) methods developed to study bound states in QED and QCD~\cite{caswell,bbl,lmr,ps}. The expansion parameter of the effective theory is the velocity $v \ll 1$ of the electron in the Ps bound state.  [We will use the velocity power counting of Refs.~\cite{lmr,lms} and treat $\alpha \sim v$.] We show that  the Ore-Powell calculation can be reconciled with Low's theorem if the latter is carefully applied considering all the energy scales present in the problem.

The three important scales for the photon decay spectrum are the electron mass $m$, the binding energy of order $m \alpha^2$, and the hyperfine splitting between the singlet and triplet states of order $m \alpha^4$. The Ore-Powell computation is valid for photon energies $E_\gamma \gg m \alpha^2$. When $E_\gamma$ is of order $m\alpha^2$, the Ps binding energy can not be neglected. The decay amplitude depends on a sum over an infinite set of excited Ps states, and can be written in terms of a Ps structure function.  We will compute the Ps structure function using NRQED to leading order in the $v$ expansion. When $E_\gamma$ is of order $m \alpha^4$, the decay amplitude is dominated by the o-Ps $\to$ p-Ps transition. We will show that the photon spectrum crosses over from an $E_\gamma$ behavior above the structure function region to an $E_\gamma^3$ behavior below the p-Ps resonance region.

The outline of the paper is as follows. In section~\ref{sec:Low} we review Low's theorem for soft photon emission and its implications for Ps decay. In section~\ref{sec:ortho}, we summarize the standard Ore-Powell computation of the decay spectrum. The NREFT calculation is discussed in the next three sections.  Section~\ref{sec:effec} discusses the matching between QED and NRQED,  section~\ref{sec:denomi} computes the decay amplitude in NRQED to leading order in $v$ in terms of Ps matrix elements, and section~\ref{sec:matrix} discusses the computation of the required matrix elements. Results and conclusions are given in Section~\ref{sec:spectrum}.

\section{Low's theorem and o-Ps $\to\;3\gamma$ decay}\label{sec:Low}

Low's theorem~\cite{low} gives the amplitude for soft photon emission in the scattering of charged particles. It states that the first two terms of the series expansion in powers of the photon energy of a radiative amplitude $X \to Y \gamma$ may be obtained  from a knowledge of the corresponding nonradiative amplitude $X \to Y$. Expanding the radiative amplitude ${\cal M}^{\mu}$ in the limit  $k\to0$, 
\begin{equation}
\epsilon_{\mu}{\cal M}^{\mu}=\frac{{\cal M}_0}{k}+{\cal M}_1+{\cal O}(k),
\label{2.1}
\end{equation}
where $\epsilon$ is the photon polarization, one finds that ${\cal M}_0$ and ${\cal M}_1$ are independent of $k$ and completely determined from the nonradiative amplitude $T_0$, its derivatives in physically allowed kinematic directions, and the electromagnetic properties of the particles involved~\cite{low}. The terms ${\cal M}_{0,1}$ are given by pole diagrams. The ${\cal O}(k)$ amplitude has both pole and non-pole contributions.

The term ${\cal M}_0$ arises from the emission of a photon by ingoing or outgoing charged particles and is proportional to $T_0$ times the universal factor  $-Q_i\, \epsilon\cdot p_i/k\cdot p_i$, summed for all the external lines in the diagram.  Note that if the nonradiative process involves no moving charged particles or is forbidden due to some selection rule, then ${\cal M}_0$ is identically zero. The term ${\cal M}_1$ can be expressed as a function of the magnetic moments, the amplitude $T_0$ and its derivatives with respect  to internal variables which are not subject to constraint, e.g.\ the energy and angles. The crucial observation is that unphysical derivatives with respect to masses ($p_i^2$)  can be shown to cancel out~\cite{low}. 

Combining the amplitude behavior with that of the phase-space, the low-frequency form of the photon spectrum is
\begin{equation}
 \frac{{\rm d}\Gamma}{{\rm d}E_\gamma}=\frac{A}{E_\gamma}+B+{\cal O}(E_\gamma),
 \label{2.2}
\end{equation}
where $A$ is proportional to $\abs{{\cal M}_0}^2$ and $B$ is the ${\cal M}_0 {\cal M}_1$ interference term. If ${\cal M}_0$ vanishes, the soft photon decay spectrum is of order $E_\gamma\,{\rm d}E_\gamma$; if both ${\cal M}_{0}$ and  ${\cal M}_{1}$vanish, it is of order $E_\gamma^3\,{\rm d}E_\gamma$.

In the three-photon decay of o-Ps, one of the photons can have an  arbitrarily small energy. The process can be then viewed as the radiative version of the o-Ps$\to 2\gamma$ decay. As the two-photon decay of o-Ps is not allowed by charge conjugation invariance, the direct application of Low's theorem yields  ${\cal M}_{0,1}=0$ so that the o-Ps$\to 3\gamma$ amplitude is of order ${\cal O}(E_\gamma)$, and the decay spectrum is
\begin{equation}
\frac{{\rm d}\Gamma_{{\rm oPs}\to 3\gamma}}{{\rm d}E_\gamma}\sim E_\gamma^3
\label{2.3}
\end{equation} 
as $E_\gamma\to 0$. 
This is in contradiction with the Ore-Powell spectrum (see Eq.~(\ref{3.14})) which vanishes linearly with $E_\gamma$, and is the contradiction pointed out in Ref.~\cite{pestieau}.

To understand the origin of the contradiction, it is worth noting that in the derivation of Low's theorem, one takes the limit $E_\gamma \to 0$ and neglects all states other than those degenerate with the incoming and outgoing states, i.e.\ one uses $E_\gamma \ll \Delta E$, where $\Delta E$ is the energy gap to excited states. The amplitudes ${\cal M}_{0,1}$ depend on the charge and magnetic moment couplings between all states degenerate with the initial or final states. A more general version of Low's theorem gives the decay spectrum for small $E_\gamma$ without taking the strict $E_\gamma \to 0$ limit. One treats all states with $\Delta E \ll E_\gamma$ as degenerate states, and includes them in the computation of charge and magnetic moment matrix elements for the purposes of Low's result. Which states are included in Low's theorem then depends on the magnitude of $E_\gamma$.

In the case of Ps decay, consider the case where the photon energy is much larger than the binding energy. Then all Ps states (including o-Ps, p-Ps, radial excitations, etc.) are degenerate for the purposes of Low's theorem. In this case, there is a non-zero magnetic dipole matrix element between o-Ps and p-Ps, so that ${\cal M}_1$ does not vanish in this extended space of states. As a result, the decay spectrum vanishes linearly with $E_\gamma$. This is the approximation under which the Ore-Powell calculation is valid, as we will see in more detail later. For energies much smaller than the o-Ps--p-Ps hyperfine splitting, p-Ps as well as radial excitations are treated as excited states, the matrix element ${\cal M}_1$ vanishes, and the spectrum is of order $E_\gamma^3$.

\section{Orthopositronium Decay Amplitude}\label{sec:ortho}

\begin{figure}
\begin{center}
\includegraphics[width=4cm]{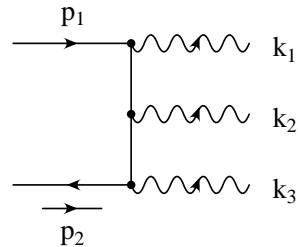}
\end{center}
\caption{Three photon annihilation graph. The graph is summed over the $3!$ permutations of the photons.\label{fig:3gamma}}
\end{figure}

The Ore-Powell calculation~\cite{OrePowell} of the $3\gamma$ annihilation of
o-Ps is given in many textbooks. We briefly review the derivation here,  using the notation of Ref.~\cite{itzykson}. The amplitude for an electron and positron to annihilate into three photons is given, to lowest order in $\alpha$, by the graph in Fig.~\ref{fig:3gamma}.
In the center-of-mass frame, the momentum of the electron is $\mathbf{p}$ and that of the positron is $-\mathbf{p}$. The on-shell decay amplitude is a function of $\mathbf{p}$.
In the NRQED power counting $\mathbf{p}$ is of order $v$, so one can expand the amplitude in a power series in $v$. The leading term in the expansion, which is all that is required here, is the $\mathbf{p} \to 0$ limit of the annihilation amplitude.

Pick a gauge where the photon polarizations $\epsilon_i$ are purely transverse, 
$\epsilon_i \cdot k_i=0$, $\epsilon^0_i=0$. The sum of all six QED graphs gives
\begin{eqnarray}
A &=&  -{ ie^3 \over 2 m^2} \chi^\dagger \Bigl[ \bm{\sigma} \cdot \bm{V} \Bigr] \phi ,\nn
\bm{V} 
&=&   \left( {\bm{\epsilon}}_2 \cdot \bm{\epsilon}_3 \right) {\bm{\epsilon}}_1 + \left( {\bm{\epsilon}}_1 \cdot \bm{\epsilon}_3 \right) {\bm{\epsilon}}_2 + \left( {\bm{\epsilon}}_1 \cdot {\bm{\epsilon}}_2 \right) \bm{\epsilon}_3\nn
&& + \left( {\bm{\delta}}_2 \cdot \bm{\epsilon}_3 \right) {\bm{\delta}}_1 + \left( {\bm{\delta}}_1 \cdot \bm{\epsilon}_3 \right) {\bm{\delta}}_2 - \left( {\bm{\delta}}_1 \cdot {\bm{\delta}}_2 \right) \bm{\epsilon}_3 \nn
&& + \left( {\bm{\delta}}_3 \cdot \bm{\epsilon}_1 \right) {\bm{\delta}}_2 + \left( {\bm{\delta}}_2 \cdot \bm{\epsilon}_1 \right) {\bm{\delta}}_3 - \left( {\bm{\delta}}_2 \cdot {\bm{\delta}}_3 \right) \bm{\epsilon}_1 \nn
&& + \left( {\bm{\delta}}_1 \cdot \bm{\epsilon}_2 \right) {\bm{\delta}}_3 + \left( {\bm{\delta}}_3 \cdot \bm{\epsilon}_2 \right) {\bm{\delta}}_1 - \left( {\bm{\delta}}_3 \cdot {\bm{\delta}}_1 \right) \bm{\epsilon}_2 , \nn
\label{3.01}
\end{eqnarray}
where 
\begin{eqnarray}
{\bm{ \delta }}_j &=& \hat {\bm{k}}_j \times  {\bm{\epsilon}}_j,\qquad
\hat {\bm{k}}_j\,= \,\frac{{\bm{k}}_j}{E_j},
\label{3.02}
\end{eqnarray}
and we use non relativistic Pauli spinors $\phi\,,\chi$ for the electron and
positron respectively,
\begin{eqnarray}
u = \left( \begin{array}{c} \phi \\ 0 \end{array} \right)\;\;\;,
\;\;\;
v = \left( \begin{array}{c} 0 \\ \chi  \end{array} \right).
\label{3.03}
\end{eqnarray}

Let us take $k_3$ as the photon with vanishing energy.  As $E_3 \to 0$, the
high-energy photons become  antiparallel, so we have $\hat{ \bm{k}}_1 =  - \hat {\bm{k}}_2$.  In this limit, Eq.~(\ref{3.01}) becomes
\begin{eqnarray}
\bm{V} &=& 2 \left( {\bm{\epsilon}}_2 \cdot \bm{\epsilon}_3 \right) {\bm{\epsilon}}_1 + 2 \left( {\bm{\epsilon}}_1 \cdot \bm{\epsilon}_3 \right) {\bm{\epsilon}}_2   \nn
&&+ 2 \left( {\bm{\epsilon}}_1 \cdot \bm{\epsilon}_2 \right)\left( \hat {\bm{k}}_1
\cdot   {\bm{\epsilon}}_3  \right) \hat {\bm{k}}_1\nn
&&+  2 \left[ \left(\hat {\bm{k}}_1 \cdot{\bm{\epsilon}}_3 \right)\left(\bm{\epsilon}_2\cdot \hat {\bm{k}}_3 \right) -  \left( \hat {\bm{k}}_1 \cdot \hat {\bm{k}}_3  \right)  \left( \bm{\epsilon}_2 \cdot {\bm{\epsilon}}_3 \right)\right] {\bm{\epsilon}}_1 \nn
&& +  2 \left[ \left( \hat {\bm{k}}_3 \cdot \hat {\bm{k}}_1 \right) \left(
{\bm{\epsilon}}_3 \cdot {\bm{\epsilon}}_1 \right) -  \left( \hat {\bm{k}}_3\cdot
{\bm{\epsilon}}_1 \right) \left(  {\bm{\epsilon}}_3 \cdot \hat {\bm{k}}_1 \right)\right] \bm{\epsilon}_2\nn
&& + 2 \left[ \left(  {\bm{\epsilon}}_1 \cdot\hat {\bm{k}}_3 \right) \left(
\bm{\epsilon}_2 \cdot {\bm{\epsilon}}_3 \right)   -  \left( {\bm{\epsilon}}_1
\cdot{\bm{\epsilon}}_3 \right) \left(  \bm{\epsilon}_2 \cdot \hat {\bm{k}}_3
\right) \right]  \hat {\bm{k}}_1, \nn
\label{3.04}
\end{eqnarray}
since ${ \bm{\epsilon}}_{1,2} \cdot \hat{\bm  {k} }_1=0$, and we have used
the identity
\begin{eqnarray}
&& \left[ \left( \bm{a}  \times \bm{b} \right) \cdot \bm{c} \right]
\left[ \left( \bm{x}  \times \bm{y} \right) \cdot \bm{z} \right] \nn
&=& \left( \bm{a} \cdot \bm{x} \right) \left( \bm{b} \cdot \bm{y} \right) \left( \bm{c} \cdot \bm{z} \right) 
+  \left( \bm{a} \cdot \bm{y} \right) \left( \bm{b} \cdot \bm{z} \right) \left( \bm{c} \cdot \bm{x} \right)\nn 
&& +  \left( \bm{a} \cdot \bm{z} \right) \left( \bm{b} \cdot \bm{x} \right) \left( \bm{c} \cdot \bm{y} \right) - \left( \bm{a} \cdot \bm{z} \right) \left( \bm{b} \cdot \bm{y} \right) \left( \bm{c} \cdot \bm{x} \right) \nn
&& - \left( \bm{a} \cdot \bm{y} \right) \left( \bm{b} \cdot \bm{x} \right) \left( \bm{c} \cdot \bm{z} \right)
- \left( \bm{a} \cdot \bm{x} \right) \left( \bm{b} \cdot \bm{z} \right) \left( \bm{c} \cdot \bm{y} \right) .\nn
\label{3.04b}
\end{eqnarray}

\subsection{Ore-Powell Decay Spectrum}

The o-Ps annihilation amplitude can be obtained from the free-particle decay amplitude in Eq.~(\ref{3.01}) by taking the matrix element between the o-Ps state and the vacuum. If the o-Ps momentum space wavefunction is $\phi_o(\mathbf{p})$ and the annihilation amplitude is $A(\mathbf{p})$, the bound-state decay amplitude is
\begin{eqnarray}
\int {{\rm d}^3 {\mathbf p} \over (2 \pi)^3} A(\mathbf{p}) \phi_o(\mathbf{p}).
\label{3.04f}
\end{eqnarray}
To lowest order in $v$, $A(\mathbf{p})$ is a constant, $A(0)$, and Eq.~(\ref{3.04f}) becomes
\begin{eqnarray}
A(0) \psi_o(0),
\label{3.04g}
\end{eqnarray}
where $\psi_o(0)$ is the o-Ps position space wavefunction at the origin, $\mathbf{x}=0$.
Using
\begin{eqnarray}
\psi_o(\mathbf{x})=\frac{1}{\left(\pi a^3\right)^{1/2}}e^{-\abs{\mathbf{x}}/a} ,\qquad a=\frac{2}{m\alpha},
\label{3.03b}
\end{eqnarray}
for the 1S wavefunction, spin-averaging $\abs{A}^2$ in Eq.~(\ref{3.01}) (see Ref.~\cite{itzykson}) gives,
\begin{eqnarray}
\frac 13\sum_{\text{spin}}\abs{A}^2 &=& {4 e^6\over 3 m^4} 
\,|\psi_o(0)|^2\sum_{\text{cyclic}} \left(1- \cos \theta_{12} \right)^2,
\label{3.10}
\end{eqnarray}
where $\cos \theta_{12}=\hat {\bm{k}}_1\cdot\hat {\bm{k}}_2$. In terms
of the dimensionless variables $x_i=E_i/m$,
\begin{eqnarray}
\left(1- \cos \theta_{13} \right) &=& 2 {x_1 + x_3 -1  \over x_1 x_3} ,
\label{3.11}
\end{eqnarray}
and the differential decay rate is
\begin{eqnarray}
{\rm d}\Gamma_{3\gamma}   &=&  {2 m \alpha^6 \over 9 \pi} {\rm d} x_1 {\rm d} x_3
\Biggl[ {(x_1 + x _3 -1 )^2 \over x_1^2 x_3^2} \nn
 &&  + {(1-x_1 )^2 \over x_3^2 (2-x_1-x_3)^2}  + {(1-x_3 )^2 \over  x_1^2(2-x_1-x_3)^2}
 \Biggr],\nn
\label{1.26}
\end{eqnarray}
using the phase space factor
\begin{eqnarray}
\frac{1}{6}{2(2m)^2 \over 256\pi^3}\,\text{d}x_1\text{d}x_3 .
\label{6.16}
\end{eqnarray}
In addition to the usual three-body phase space, we have $1/6$ for three identical particles in the final state, and $2(2m)$ to convert from the non-relativistic normalization of the o-Ps state to the relativistic normalization.

\begin{figure}
\begin{center}
\includegraphics[width=8cm]{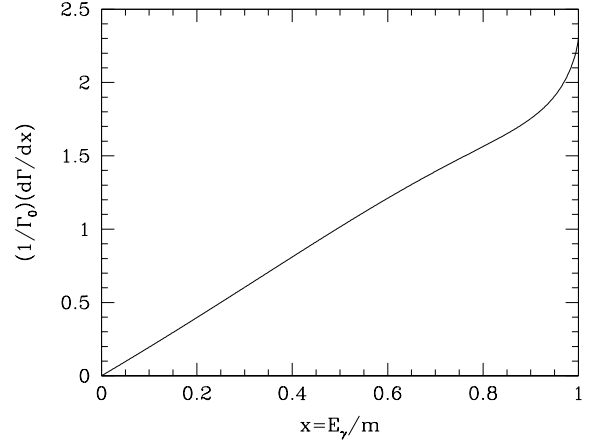}
\end{center}
\caption{The Ore-Powell orthopositronium decay spectrum.\label{fig:spec}}
\end{figure}

For the decay spectrum we can perform the integration over $x_1$ to obtain 
($x\equiv x_3$)
\begin{eqnarray}
{{\rm d}\Gamma_{3\gamma} \over {\rm d} x}&=&{4 m \alpha^6 \over 9 \pi} 
\Biggl[\frac{2 - x}{x} 
+ \frac{\left( 1 - x\right) \,x}{{\left( 2 - x \right) }^2} 
- \frac{2\,{\left( 1 - x \right) }^2\,\log (1 - x)}{{\left( 2 - x \right)}^3}\nn
&& + \frac{2\,\left( 1 - x\right) \,\log (1 - x)}{x^2} \Biggr],
\label{3.13}
\end{eqnarray}
which is the Ore-Powell result~\cite{OrePowell}, and is plotted in Fig.~\ref{fig:spec}.
Integrating Eq.~(\ref{3.13}) over $x$ gives the total decay rate
\begin{eqnarray}
\Gamma_{3\gamma}  &=&{2 \left(\pi^2-9\right) m \alpha^6 \over 9 \pi}.
 \label{1.23}
\end{eqnarray}
The low-energy photon spectrum results from the $x \to 0$ limit of Eq.~(\ref{3.13}),
\begin{eqnarray}
{{\rm d}\Gamma_{3\gamma} \over {\rm d} x}  &=&  
{2 m \alpha^6 \over 9 \pi} \left[ {5 x \over 3} + \mathcal{O}(x^2) \right],
\label{3.14}
\end{eqnarray}
which vanishes linearly with the energy of the radiated photon.
 
\section{Effective Theory Computation: Matching Condition}\label{sec:effec}

The NREFT computation provides a systematic way of including bound state effects in the computation of the o-Ps decay amplitude. One computes the decay amplitude in QED. The NRQED Hamiltonian is constructed to reproduce the same amplitude, neglecting any binding effects. Once the NRQED Hamiltonian has been determined, it can be used to compute the decay including binding corrections.

The bound state dynamics is described by the Coulomb Hamiltonian for an $e^+e^-$ system in interaction with the quantized electromagnetic field:
\begin{eqnarray}
H &=& H_0+H_{\text{int}} \nn
H_0 &=& \frac{\mathbf{p}^2}{m}-\frac{\alpha}{r} \nn 
H_{\text{int}} &=& -\mu\, [{\bm{\sigma}_{\phi}+\bm{\sigma}_{\chi}}]
\cdot{\mathbf{B}}
-e\,{\mathbf{x}}\cdot{\mathbf{E}}
\label{5.01}
\end{eqnarray}
with $\mathbf{x}=\mathbf{x_1}-\mathbf{x_2}$ the relative position of the pair of leptons, $\mathbf{p}_1=\mathbf{p}$, $\mathbf{p}_2=-\mathbf{p}$, and $\bm{\sigma}_{\phi},\,\bm{\sigma}_{\chi}$ the Pauli matrices acting on the electron and positron spinors ($\mu=e/2m$).

The interaction Hamiltonian $H_{\text{int}}$ has been written in gauge invariant form in terms of the electric and magnetic fields. It can be obtained by a G\"oppert-Mayer
transformation of the usual Hamiltonian with a ${\mathbf{p}}- e {\mathbf{A}}$ coupling (see e.g.~Ref.~\cite{cohen,pinedasoto}).  The G\"oppert-Mayer transformation removes the vector potential coupling in the covariant derivative, so that the electron and positron behave as neutral particles. The electromagnetic interactions are through multipole interactions with the electric and magnetic fields. The electric and magnetic dipole interactions are shown in Eq.~(\ref{5.01}). The higher multipoles are higher order in $v$.
Gauge invariance, and the computation using $\mathbf{p} \cdot \mathbf{A}$ interactions is discussed in Appendix.~\ref{sec:gauge}.

The electric and magnetic fields $\mathbf{E},\, \mathbf{B}$  in $H_{\text{int}}$ are evaluated
in the dipole approximation, which is valid when the spatial extent of the positronium state,
$a=2a_0=2/m\alpha$, is small with respect the wavelength of the radiated photon (i.e. when
$k\ll m\alpha$). The low energy photon is described by an ultrasoft field in NRQED, and the
multipole expansion follows from the NRQED velocity expansion~\cite{multipole,lmr}. Our NRQED
computation is valid provided the soft photon energy is small compared with $m \alpha$. The
NRQED result is valid even though the other two photons in the decay have energies of order
$m$, because the annihilation into hard photons is a short distance coefficient in the NRQED
Hamiltonian. It is possible to extend the NRQED computation to photon energies which are small
compared to $m$ but not $m\alpha$ by not using the multipole expansion.

We shall use time-ordered (or ``old-fashioned") perturbation theory (TOPT) in the effective theory calculations, as it is more suitable for non-relativistic interactions. 
We recall that  in TOPT vertices conserve only three-momenta and the virtual states are 
always on-shell. It is the violation  of energy that characterizes the intermediate states 
rather than the  off-shellness of the particles, as in covariant perturbation theory. 

The Coulomb Hamiltonian $H_0$ is the leading term in the velocity power counting. The kinetic energy and Coulomb potential are the same order in $v$. The energies and wavefunctions of $H_0$ are thus the Coulomb wavefunctions with reduced mass $m/2$.
The electric and magnetic dipole interaction terms are treated as perturbations. 
The long distance part of the o-Ps decay amplitude involves these interactions.

The p-Ps decay amplitude has no long-distance contribution, since both photons are hard, 
with energy $m$. The p-Ps annihilation amplitude is purely short-distance, and is given 
by computing the on-shell $e^+e^-$ annihilation amplitude in QED as a power series in the 
lepton momentum $\mathbf{p}$. For this paper, we need the first two terms in the $v$ expansion,
 which are given in Appendix.~\ref{sec:para}. Since the p-Ps decay amplitude is purely 
 short distance, it can be written as a local operator in NRQED, with the hard photons 
 treated as external sources from the point of view of the effective theory. The photons
  carry away the energy of order $m$ produced in the annihilation, so the NRQED operator
   only depends on the scale $m\alpha$. The first two terms in the $v$ expansion 
   computed in Appendix~\ref{sec:para} give the annihilation operators 
   in the Lagrangian (a sum on $\mathbf{p}$ is implicit)
 \begin{eqnarray}
- i{e^2 \over 16 m^3}  \epsilon_{\mu \nu \alpha \beta} {\cal F}^{\mu \nu} {\cal F}^{\alpha \beta}
\chi^\dagger_{-\mathbf{p}} \phi_{\mathbf{p}} +{e^2 \over 2m^4} {\cal F}^i{}_\alpha {\cal F}^{j \alpha}\mathbf{p}^i 
\chi^\dagger_{-\mathbf{p}} \bm{\sigma}^j \phi_{\mathbf{p}}\nn
\label{annop}
\end{eqnarray}
where $\psi_{\mathbf{p}}$ annihilates electrons with momentum ${\mathbf{p}}$ and 
$\chi^\dagger_{-\mathbf{p}}$ annihilates positrons with 
momentum $-{\mathbf{p}}$. We use the convention $\epsilon_{0123}=+1$ for the
Levi-Civita tensor. The field strength tensors of the hard photons have been denoted by ${\cal F}$ to emphasize that they are external sources, and not dynamical fields in NRQED. The operators Eq.~(\ref{annop}) allow one to compute the annihilation amplitude including all angular and polarization dependence, including that of the hard photons. One can instead integrate out the external sources ${\cal F}$ to get local four-fermion annihilation operators. These allow one to compute the angular and polarization dependence on the soft photon after averaging over all possible states of the hard photons.
The four-fermion operators we will need are the S- and P-wave annihilation operators in Ref.~\cite{bbl},
\begin{eqnarray}
L &=& {f_{\gamma\gamma}({}^1S_0)\over m^2} {\cal O}({}^1S_0) + {f_{\gamma\gamma}({}^3P_0) \over m^4} {\cal O}({}^3P_0)+ {f_{\gamma\gamma}({}^3P_2)\over m^4} {\cal O}({}^3P_2) \nn 
\label{annlag}
\end{eqnarray}
where the operators are~\cite{bbl}
\begin{eqnarray}
{\cal O}({}^1S_0) &=& \chi^\dagger_{-\mathbf{p}} \phi_{\mathbf{p}} \ket{0} \bra{0} \phi^\dagger_{\mathbf{p}}  \chi_{-\mathbf{p}} \nn
{\cal O}({}^3P_0) &=& \frac 1 3 \chi^\dagger_{-\mathbf{p}} (\mathbf{p} \cdot \bm{\sigma} )\phi_{\mathbf{p}} \ket{0} \bra{0} \phi^\dagger_{\mathbf{p}}  ( \mathbf{p} \cdot \bm{\sigma}) \chi_{-\mathbf{p}} \nn
{\cal O}({}^3P_2) &=& \chi^\dagger_{-\mathbf{p}} (\mathbf{p}^{(i} \bm{\sigma} ^{j)} )\phi_{\mathbf{p}} \ket{0} \bra{0} \phi^\dagger_{\mathbf{p}} (  \mathbf{p}^{(i} \bm{\sigma} ^{j)})  \chi_{-\mathbf{p}} \nn
\end{eqnarray}
and the coefficients are given in Appendix~A of Ref.~\cite{bbl}
\begin{eqnarray}
\text{Im} f_{\gamma\gamma}({}^1S_0) &=& \pi \alpha^2, \nn
\text{Im} f_{\gamma\gamma}({}^3 P_0) &=& 3 \pi \alpha^2, \nn
\text{Im} f_{\gamma\gamma}({}^3 P_2) &=& {4 \pi \alpha^2 \over 5}.
\label{anncoeff}
\end{eqnarray}
The annihilation Lagrangian can only be used to compute the square of the decay amplitude.

 The o-Ps decay amplitude has both short distance and long-distance contributions.
The matching condition for the o-Ps decay amplitude in the effective theory is given by computing the difference between the QED o-Ps decay amplitude in the soft photon limit, Eq.~(\ref{3.04}), and the corresponding amplitude computed in the effective theory. The difference between the two computations gives the total short-distance o-Ps annihilation contribution in the effective Hamiltonian. We will compute the matching conditions using the annihilation amplitude Eq.~(\ref{annop}), and use it to compute the decay spectrum. In Sec.~\ref{sec:annresult} we show how the decay spectrum can be obtained using Eq.~(\ref{annlag}).

\subsection{Magnetic Dipole}

\begin{figure}
\begin{center}
\includegraphics[width=4cm]{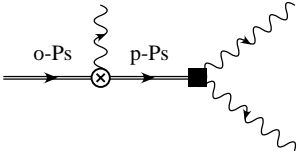}
\caption{Magnetic dipole graph for o-Ps annihilation. The solid square denotes the  p-Ps annihilation vertex.\label{fig:magnetic}}
\end{center}
\end{figure}

The magnetic term in $H_{\text{int}}$ can induce a $1\,^3S_1\to 1\,^1S_0$ transition between Coulomb $e^+e^-$ states.  The effective theory graph for magnetic dipole emission is shown in Fig.~\ref{fig:magnetic}, and contributes to the long-distance o-Ps decay amplitude.
The amplitude is:
\begin{eqnarray}
A_m&=&\sum_n{ i \me{0}{ A^{(2\gamma)}}{n}
\me{n}{i\,\mu \,[{\bm{\sigma}_{\phi}+\bm{\sigma}_{\chi}}]
\cdot{\mathbf{B}}}{\text{o-Ps}}\over E_o-E_n-E_3} \nn
&=& i\,{ \me{0}{ A^{(2\gamma)}}
{1\,^1S_0}\me{1\,^1S_0}{i\,\mu \,[{\bm{\sigma}_{\phi}+\bm{\sigma}_{\chi}}]
\cdot{\mathbf{B}}}{\text{o-Ps}}
\over E_o-E_p-E_3}\nn
\label{5.11}
\end{eqnarray}
where the only allowed intermediate state in the dipole approximation is the p-Ps ground state, with energy $E_p$, so that $E_o-E_p=\Delta E_{\,\text{hfs}}$. The  energy of the emitted photon is $E_3$. Recoil effects in the energy propagator are suppressed by $E_3^2/m$ and are higher order in $v$. 

The magnetic dipole amplitude above is easily evaluated 
\begin{eqnarray}
\me{1\,^1S_0}{i\,\mu \,[{\bm{\sigma}_{\phi}+\bm{\sigma}_{\chi}}]
\cdot{\mathbf{B}}}{\text{o-Ps}}&=&{e \over \sqrt{2}m}E_3\,
\bm{\delta}_3 \cdot \chi^{\dagger}\,\bm{\sigma}\,\phi .\nn
\label{5.12}
\end{eqnarray}
The quantity $A^{(2\gamma)}$ is the $e^+e^-\to 2\gamma$ amplitude with 
fermions at rest calculated in Appendix~\ref{sec:para}, see Eq.~\ref{A7}. 
For $e^+e^-$ $S$-wave states only the constant part of the $A^{(2\gamma)}$ amplitude ($W_0$) contributes. When projected in the spin-singlet configuration ($\chi^{\dagger}\psi\to\sqrt{2}\,$), it yields
\begin{eqnarray}
\me{0}{ A^{(2\gamma)}}
{1\,^1S_0}&=& {e^2 \over \sqrt{2} m} 
\left[ \bm{\delta}_1 \cdot \bm{\epsilon}_2 + \bm{\delta}_2 \cdot \bm{\epsilon}_1  \right]
\psi_o(0) . \nn
\label{5.13}
\end{eqnarray}

To compute the o-Ps decay amplitude matching coefficient, one compares the effective theory result with the QED computation which neglects bound state effects. Neglecting
 the hyperfine mass difference in the energy denominator of Eq.~(\ref{5.12}) gives
\begin{eqnarray}
A_m&=& -{ie^3 \over 2 m^2}\,\psi_o(0)\, \chi^{\dagger}
\bm{\delta}_3 \cdot \bm{\sigma}
\left[ \bm{\delta}_1 \cdot \bm{\epsilon}_2 + \bm{\delta}_2 \cdot \bm{\epsilon}_1  \right] 
\phi , \nn
\label{5.14}
\end{eqnarray}
so we can define, in analogy with Eq.~(\ref{3.01}),
\begin{eqnarray}
\bm{V}_m &=&  \bm{\delta}_3  \left[ \bm{\delta}_1 \cdot \bm{\epsilon}_2 + \bm{\delta}_2 \cdot \bm{\epsilon}_1  \right] .
\label{5.15}
\end{eqnarray}
The limit of $\bm{V}_m$ as $E_3\to 0$ is 
\begin{eqnarray} 
\bm{V}_m &=&2 \left[ \left(\hat {\bm{k}}_1 \cdot{\bm{\epsilon}}_3 \right)  \left(  \bm{\epsilon}_2\cdot \hat {\bm{k}}_3 \right) -  \left( \hat {\bm{k}}_1 \cdot  \hat {\bm{k}}_3  \right)  \left( \bm{\epsilon}_2 \cdot {\bm{\epsilon}}_3 \right)\right] {\bm{\epsilon}}_1 \nn
&& +  2 \left[ \left( \hat {\bm{k}}_3 \cdot \hat {\bm{k}}_1 \right) \left(  {\bm{\epsilon}}_3 \cdot {\bm{\epsilon}}_1 \right) -  \left( \hat {\bm{k}}_3 \cdot {\bm{\epsilon}}_1 \right) \left(  {\bm{\epsilon}}_3 \cdot \hat {\bm{k}}_1 \right)\right] \bm{\epsilon}_2\nn
&& + 2 \left[ \left(  {\bm{\epsilon}}_1 \cdot\hat {\bm{k}}_3 \right) \left( \bm{\epsilon}_2 \cdot {\bm{\epsilon}}_3 \right)   -  \left( {\bm{\epsilon}}_1 \cdot{\bm{\epsilon}}_3 \right) \left(  \bm{\epsilon}_2 \cdot \hat {\bm{k}}_3 \right) \right]  \hat {\bm{k}}_1 .\nn
\label{5.16}
\end{eqnarray}

\subsection{Electric Dipole}

\begin{figure}
\begin{center}
\includegraphics[width=4cm]{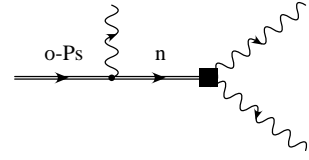}
\end{center}
\caption{Electric dipole transition for o-Ps decay. The solid square denotes the p-Ps annihilation vertex.\label{fig:electric}}
\end{figure}

The electric dipole term $-e\,{\mathbf{x}}\cdot{\mathbf{E}}$ in $H_{\text{int}}$
can change orbital angular momentum by one unit, allowing for transitions from ground state
 o-Ps to $n\,^3P_{0,2}$ states ($n\ne 1$). We do not consider
 intermediate $n\,^3P_1$ states as they cannot decay to
 two photons~\cite{yang}.  The effective theory graph for electric dipole emission is shown in Fig.~\ref{fig:electric},
with amplitude
\begin{eqnarray}
A_e&=&\sum_n { i \me{0}{ A^{(2\gamma)}}{n}
\me{n}{i\,e \mathbf{x} \cdot \mathbf{E}}{\text{o-Ps}}
\over E_o-E_n-E_3},
\label{5.21}
\end{eqnarray}
and
\begin{eqnarray}
\me{n}{i\,e \mathbf{x} \cdot \mathbf{E}}{\text{o-Ps}}&=&e\,E_3
\me{n}{\mathbf{x}\cdot \bm{\epsilon}_3}{\text{o-Ps}} .
\label{5.22}
\end{eqnarray}
For the matching computation, the energy separation between o-Ps and the intermediate states, $E_o-E_n\sim {\cal O}(m\alpha^2)$ is taken to be much smaller than the photon energy $E_3$. In this case
\begin{eqnarray}
A_e&=&-ie \sum_n  \me{0}{ A^{(2\gamma)}}{n}
\me{n}{ \mathbf{x} \cdot \bm{\epsilon}_3 }{\text{o-Ps}}\nn
&=&-i e   \me{0}{ A^{(2\gamma)}  \mathbf{x} \cdot \bm{\epsilon}_3 }{\text{o-Ps}}.
\label{5.23}
\end{eqnarray}
The two-photon amplitude $A^{(2\gamma)}$ has terms which are a constant or 
linear in $\mathbf{p}$. The constant term $W_0$ yields the matrix element
\begin{eqnarray}
\me{0}  {\mathbf{x} }{\text{o-Ps}} = 0
\label{5.24}
\end{eqnarray}
since $\mathbf{x} \,\psi_o(\mathbf{x})$ vanishes at the origin. 
The linear terms $\mathbf{W}_1 \cdot \mathbf{p}$ give
\begin{eqnarray}
\me{0}  {\mathbf{W}_1\cdot \mathbf{p} \,
\mathbf{x} \cdot \bm{\epsilon}_3 }{\text{o-Ps}},
\label{5.25}
\end{eqnarray}
which can be evaluated using
\begin{eqnarray}
\mathbf{p}^i \mathbf{x}^j& =& \frac 1 3 \delta^{ij}  \mathbf{p} \cdot \mathbf{x}
\,+\,\frac 12  \left( \mathbf{p}^i \mathbf{x}^j - \mathbf{p}^j \mathbf{x}^i \right) \nn
&&+\,\frac 12 \left( \mathbf{p}^i \mathbf{x}^j + \mathbf{p}^j \mathbf{x}^i 
-\frac 2 3 \delta^{ij} \mathbf{p} \cdot \mathbf{x} \right).
\label{5.26}
\end{eqnarray}
The o-Ps state is an $S$ state, so to get a non-zero matrix element, the operator must
have zero angular momentum. Only the first term contributes, and the $\mathbf{p}$ must 
act on $\mathbf{x}$, otherwise the wavefunction vanishes at the origin. This gives
\begin{eqnarray}
\me{0}  {\mathbf{W}_1\cdot \mathbf{p} \,\mathbf{x} \cdot \bm{\epsilon}_3 }
{\text{o-Ps}} &=&-i   \me{0}{\mathbf{W}_1\cdot \bm{\epsilon}_3 }{\text{o-Ps}}
\nn
& = &- i\,\psi_o(0)\,\chi^{\dagger}\mathbf{W}_1\cdot \bm{\epsilon}_3 \phi .
\label{5.27}
\end{eqnarray}
Substituting the $\mathbf{W}$ term from Eq.~(\ref{A11}) we get
\begin{eqnarray}
A_e&=& -{i e^3 \over m^2} \psi_o(0)\,\chi^{\dagger} 
\Bigl[\left( \bm{\epsilon}_3 \cdot \hat{\bm{k}}_1 \right)\left(\bm{\sigma} \cdot \hat{\bm{k}}_1 \right) \left(\bm{\epsilon}_1 \cdot \bm{\epsilon}_2 \right) \nn
&&+ \left( \bm{\epsilon}_3 \cdot \bm{\epsilon}_1 \right)
 \bm{\sigma} \cdot \bm{\epsilon}_2+ \left( \bm{\epsilon}_3 \cdot \bm{\epsilon}_2 \right)
 \bm{\sigma} \cdot \bm{\epsilon}_1\Bigr]\phi
 \label{5.28}
\end{eqnarray}
so that 
\begin{eqnarray}
\bm{V}_e &=&   2 \left( \bm{\epsilon}_3 \cdot \bm{\epsilon}_1 \right)
\bm{\epsilon}_2+2  \left( \bm{\epsilon}_3 \cdot \bm{\epsilon}_2 \right) \bm{\epsilon}_1 + 2 \, \hat{\bm{k}}_1 \left( \bm{\epsilon}_3 \cdot \hat{\bm{k}}_1 \right) \left(\bm{\epsilon}_1 \cdot \bm{\epsilon}_2 \right) .\nn
 \label{5.30}
\end{eqnarray}

The sum of the magnetic and electric dipole transitions in the effective theory, Eqs.~(\ref{5.16}) and (\ref{5.30}) gives the full theory amplitude Eq.~(\ref{3.04}). The matching condition, which is the difference of the two results, vanishes. Thus there is no additional three-photon annihilation term in the  NRQED Hamiltonian, and the entire soft-photon o-Ps annihilation in the effective theory is due to electric and magnetic radiation followed by two-photon annihilation of p-Ps. This must happen, because the amplitude Eq.~(\ref{3.04}) in the full theory is $\mathcal{O}(1)$ as $E_\gamma \to 0$. Any local gauge invariant operator in the NRQED Hamiltonian depends on the soft-photon field-strength tensor, and gives an amplitude that is order $\mathcal{O}(E_\gamma)$.

\section{Effective theory decay  amplitude including energy denominators}\label{sec:denomi}

The decay amplitude in the effective theory can now be computed without neglecting the
energy difference between $\text{o-Ps}$ and the intermediate states. The matching calculation of the previous section shows that the entire contribution comes from the magnetic and electric graphs in Figs.~\ref{fig:magnetic},\ref{fig:electric}.
  
The magnetic dipole amplitude including energy denominators is from Eq.~(\ref{5.11})
\begin{eqnarray}
-iA_m &=& {e^3 \over 2 m^2} \psi_o(0)\,\bm{\delta}_3 \cdot  \bm{\sigma} 
\left[ \bm{\delta}_1 \cdot \bm{\epsilon}_2 + \bm{\delta}_2 \cdot \bm{\epsilon}_1  \right]
 {  \omega\over \Delta E_{\,\text{hfs}}-\omega}\nn
\label{6.01}
\end{eqnarray}
while the electric dipole emission amplitude is
\begin{eqnarray}
-iA_e&=&e \omega \sum_n {  \me{0}{ A^{(2\gamma)}}{n}\me{n}{\mathbf{x} \cdot \bm{\epsilon}_3}{\text{o-Ps}}
\over E_o-E_n-\omega} 
\label{6.02}
\end{eqnarray}
where $\omega=E_3$ is the soft photon energy. The $\mathbf{x \cdot E}$ operator connects 
to $p$-wave states, so only the $\mathbf{W}_1$ part of
$A^{(2\gamma)}$ is relevant,
\begin{eqnarray}
-iA_e&=&e \omega \sum_n { \me{0}{ \mathbf{W}_1 \cdot \mathbf{p} }{n}
\me{n}{\mathbf{x} \cdot \bm{\epsilon}_3}{\text{o-Ps}}
\over E_o-E_n-\omega} .
\label{6.03}
\end{eqnarray}

We can work out the matrix elements above with the help of the
Wigner-Eckart theorem~\cite{wigner}. Define the reduced matrix elements
\begin{eqnarray}
\me{n,1,0}{ z  }{\text{o-Ps}} &=& \rme{n,1}{x }{\text{o-Ps}}\nn
\me{0}{  p_z }{n,1,0} &=&  \rme{0}{ p }{n,1},
\label{6.07}
\end{eqnarray}
where $\ket{n\ell m}$ are the Ps states, and only $\ell=1$ contributes.
In terms of the reduced matrix elements, Eq.~(\ref{6.03}) reads
\begin{eqnarray}
-iA_e&=& e \omega \sum_{n}{ \chi^{\dagger}\mathbf{W}_1 \cdot \bm{\epsilon}_3
\phi\over E_o-E_n-\omega} \nn
&&\times \rme{0}{ p }{n,1}\rme{n,1}{x  }{\text{o-Ps}}.
\label{6.08}
\end{eqnarray}
It is convenient to define
\begin{eqnarray}
a_m(\omega) &=&-  { \omega\over \Delta E_{\,\text{hfs}}-\omega} \nn
a_e(\omega) &=&- i {\omega \over \psi_o(0)} \sum_{n}  { \rme{0}{ p }{n,1}\rme{n,1}{x  }{\text{o-Ps}} \over E_o-E_n-\omega} 
\label{6.09}
\end{eqnarray}
so that
\begin{eqnarray}
-i A_m &=&-{e^3 \over 2m^2} \psi_o(0) a_m \,\chi^{\dagger} \,\bm{\delta}_3 \cdot  \bm{\sigma} \left[
\bm{\delta}_1 \cdot \bm{\epsilon}_2 + \bm{\delta}_2 \cdot \bm{\epsilon}_1 
\right] \phi \nn
-i A_e &=& -{e^3 \over m^2} \psi_o(0) a_e\,\chi^{\dagger}\Biggl[ 
 \left(\bm{\epsilon}_3 \cdot \hat{\bm{k}}_1 \right)\left(\bm{\sigma} \cdot \hat{\bm{k}}_1 \right) \left(\bm{\epsilon}_1 \cdot \bm{\epsilon}_2 \right) \nn
&& + \left( \bm{\epsilon}_3 \cdot \bm{\epsilon}_1 \right)
 \bm{\sigma} \cdot \bm{\epsilon}_2+ \left(\bm{\epsilon}_3 \cdot \bm{\epsilon}_2 \right)
 \bm{\sigma} \cdot \bm{\epsilon}_1\Biggr] \phi .
\label{6.10}
\end{eqnarray}
From the equations above we already see that in the limit $\omega \to 0$ the effective theory amplitude $A_m+A_e$ vanishes as $\omega$, in agreement with Low's theorem. In the energy range $\Delta E_{\,\text{hfs}}\ll\omega\ll m\alpha^ 2$ the electric dipole
amplitude still behaves as ${\cal O}(\omega)$, but the magnetic amplitude is ${\cal O}(1)$. For the region above $m\alpha^2$ both amplitudes are ${\cal O}(1)$ and the Ore-Powell behavior is restored. 

Neglecting energy denominators, (i.e.\ taking the limit $\omega \gg E_n-E_o$) the amplitudes $a_m$ and $a_e$ reduce to 
\begin{eqnarray}
a_m(\omega \to \infty) &\to& 1 \nn
a_e(\omega \to \infty) &\to&{  i \over \psi_o(0)}  \sum_{n}   \rme{0}{ p }{n,1}\rme{n,1}{x  }{\text{o-Ps}}  \nn
&=&1
\label{6.11}
\end{eqnarray}
and we recover Eqs.~(\ref{5.14}) and (\ref{5.28}).

To compute the spectrum from our effective theory calculation we first need to to sum over polarizations. A spin-1 state of o-Ps with polarization $\bm{\varepsilon}$ is obtained by the replacement
\begin{eqnarray}
\left[\, \phi\,\chi^{\dagger}\right]_{\alpha\,\beta} &\to& \frac{1}{\sqrt{2}} 
\left[\bm{\varepsilon}^* \cdot \bm{\sigma} \right]_{\alpha\,\beta} ,
\label{6.12}
\end{eqnarray}
and the matrix element of $\bm{\sigma}$ is then
\begin{eqnarray}
\langle \phi\,\bm{\sigma}\,\chi^{\dagger}\rangle_{\bm{\varepsilon}}
 &=& \frac{1}{\sqrt{2}}\,\bm{\varepsilon}^{*\,i} 
\text{Tr} \left[ \bm{\sigma}\, \bm{\sigma}^i \right]
 \,=\, \sqrt{2}\,\bm{\varepsilon}^{*} .
\label{6.13}
\end{eqnarray}
The sum over polarizations is
\begin{eqnarray}
\sum \bm{\varepsilon}_i^* \bm{\varepsilon}_j &=& \delta_{ij},\nn
\sum \bm{\epsilon}_i^* \bm{\epsilon}_j &=& \delta_{ij}-\hat{ \bm{k} }_i \hat{\bm{k}}_j,
\label{6.14}
\end{eqnarray}
for o-Ps and photons, respectively. Using these, it is straightforward to obtain the spin-averaged matrix element squared and summed over photon polarizations
\begin{eqnarray}
\lefteqn{\frac{1}{3}\sum_{\bm{\varepsilon}}\sum_{\bm{\epsilon}_i}\,\left| A_m+A_e
\right|^2=} \\
&& \frac{8}{3} {e^6 \over m^4} \abs{\psi_o(0)}^2
\left\{\abs{a_m}^2 + \abs{a_e}^2 \left(2 + \cos^2 \theta
\right)\right\}\nn
 \label{6.15}
\end{eqnarray}
where $\cos \theta = \hat {\bm{k}}_1 \cdot \hat{\bm{k}}_3$.
This agrees with Eq.~(\ref{3.10}) if one uses Eq.~(\ref{6.11}). 

The phase space factor Eq.~(\ref{6.16}) can be written in terms of the variables 
$\cos \theta$ and $x_{3}$ by using the relation (\ref{3.11}):
\begin{eqnarray}
\text{d}x_1\text{d}x_3&=&\frac{x_3}{2}{1-x_3 \over \left[1-\frac{x_3}{2}
(1-\cos \theta)\right]^2}\,{\rm d}\!\cos \theta\,{\rm d} x_3 \nn
&\simeq& \frac{x_3}{2}\,{\rm d}\!\cos \theta\,{\rm d} x_3.
\label{6.17}
\end{eqnarray}
The above approximation is valid in the endpoint region where $x_3\to 0$. The allowed region of integration for $\cos \theta$ is $-1 \le \cos \theta \le 1$.

As $a_{e,m}$ are functions only of $x_3$, the $\cos\theta$ integration can
be performed to give the differential rate (writing $x_3 \to x$):
\begin{eqnarray}
{ {\rm d}\Gamma \over  {\rm d}x }  &=&{m \alpha^6 \over 9 \pi} x \left[\abs{a_m}^2 +  \frac 7 3 \abs{a_e}^2\right] .
\label{6.18}
\end{eqnarray}
The magnetic and electric amplitudes $a_m$ and $a_e$ are computed in the next section.

\section{Computing Matrix Elements}\label{sec:matrix}

In this section, we compute the magnetic dipole amplitude $a_m$ and the electric dipole amplitude $a_e$.

\subsection{Magnetic Dipole}

For the magnetic term $a_m$, the energy denominator can vanish, so we include the widths,
\begin{eqnarray}
a_m &=& - { \omega\over \Delta E_{\,\text{hfs}} -\omega + i \gamma_{\text{op}}} ,
\label{7.01}
\end{eqnarray}
where
\begin{eqnarray}
\Delta E_{\,\text{hfs}} &=& \frac{7}{12}\, m \,\alpha^4 ,\nn
\gamma_{\text{op}} &=& \frac12\left(\Gamma_o + \Gamma_p \right) \approx \frac 1 4 m\, \alpha^5 .
\label{7.02}
\end{eqnarray}
This gives the Breit-Wigner form
\begin{eqnarray}
\abs{a_m}^2 &=&  { \omega^2 \over \left( \omega -  \Delta E_{\,\text{hfs}}  \right)^2 + \gamma_{\text{op}}^2} .
\label{7.03}
\end{eqnarray}
The magnetic amplitude behaves as
\begin{eqnarray}
a_m(\omega) &=& \left\{ \begin{array}{cl}
1 &\qquad \omega \gg  \Delta E_{\,\text{hfs}}, \\[5pt]
-{\omega \over  \Delta E_{\,\text{hfs}} } &\qquad \omega \ll  \Delta E_{\,\text{hfs}}.
 \end{array}\right.
\label{7.03a}
\end{eqnarray}

\subsection{Electric Dipole}

We can rewrite the electric dipole amplitude, Eq.~(\ref{6.09}), as
\begin{eqnarray}
a_e &=& i {\omega\over \psi_o(0)} \me{0}{ p_z \left[ \sum_{n,m} {|n,1,m\rangle \langle n,1,m| \over E_n-E_o+\omega} \right]
z}{\text{o-Ps}} \nn
\label{7.11}
\end{eqnarray}
where the term between brackets is the $p$-wave Coulomb Green's function $G_1$ defined in Appendix~\ref{sec:coulomb} and evaluated at energy $-k^2/m=E_o-\omega$. In position space, using Eq.~(\ref{C4}),
\begin{eqnarray}
\sum_{n,m} {\langle \mathbf{x}|n,1,m\rangle \langle n,1,m| \mathbf{y}\rangle
\over E_n+k^2/m}&=&
3\,(\mathbf{x}\cdot\mathbf{y})\,G_1(x,y;k),\nn
\label{7.12}
\end{eqnarray}
gives the electric dipole amplitude 
\begin{eqnarray}
a_e (\omega) &=&  {4\pi\omega\over \psi_o(0)} \int_0^{\infty} {\rm d}  y\,y^4\,
G_1(0,y;k)\,\psi_o(y).
\label{7.13}
\end{eqnarray}
The integral in Eq.~(\ref{7.13}) can be evaluated numerically as a function
of the soft photon energy $\omega$ using Eq.~(\ref{C16}) for $\ell=1$:
\begin{eqnarray}
G_1(0,y;k)&=&{m k^3 \over 3 \pi}e^{-k y }
\,\Gamma(2-\nu)\,U(2-\nu,4,2ky),\nn
\label{7.14}
\end{eqnarray}
where $U$ is a confluent hypergeometric function, and $\nu = \alpha m /(2 k)$. We note Eq.~(\ref{7.14}) for the $p$-wave Coulomb Green's function was used in Ref.~\cite{penin}, with an additional factor of 6. Because of this discrepancy, we have given a derivation of Eq.~(\ref{7.14}) in Appendix~\ref{sec:coulomb}.

The electric amplitude behaves as
\begin{eqnarray}
a_e(\omega) &=& \left\{ \begin{array}{cl}
1 & \qquad \omega \gg m \alpha^2, \\[5pt]
{ 2 \omega \over m \alpha^2 } & \qquad \omega \ll m \alpha^2.
 \end{array}\right.
\label{7.14a}
\end{eqnarray}

\subsubsection*{Note Added}

Recently, Voloshin~\cite{voloshin} has shown that Eq.~(\ref{7.13}) can be written in terms of a hypergeometric function,
\begin{eqnarray}
a_e(\omega) &=& {(1-\nu)(3+5\nu) \over 3 (1+\nu)^2} \nn
&&+ {8 \nu^2(1-\nu) \over 3(2-\nu)(1+\nu)^3}\ {}_2F_1\left(2-\nu,1,3-\nu;{\nu - 1 \over \nu+1} \right).\nn
\end{eqnarray}

\subsection{Electric Dipole: Alternative Derivation}

The electric dipole amplitude $a_e$ can be obtained in a different representation by using the explicit form of the Coulomb wavefunctions. We follow the notation of Bethe and Salpeter~\cite{bethe} and use the expressions found therein. Note that all quantities are dimensionless, with length, momentum and energy measured in units of $1/(\mu\alpha)$, $\mu \alpha$ and $\mu \alpha^2$, respectively, where $\mu=m/2$ is the reduced mass.

For the discrete wavefunctions the dipole reduced matrix elements read
\begin{eqnarray}
\rme{n,1}{x  }{\text{o-Ps}} &=&  \left[ {1 \over 3} { 2^8 n^7 (n-1)^{2n-5} \over (n+1)^{2n+5}} \right]^{1/2},
\label{8.31}
\end{eqnarray}
\begin{eqnarray}
\rme{0}{ p }{n,1} 
&=& - i  \psi_o(0){1 \over 4 \sqrt{3} } \left({2 \over n }\right)^{5/2} \left[{(n+1) (n-1)\over 2}
 \right]^{1/2}.\nn
\label{8.35}
\end{eqnarray}
Define
\begin{eqnarray}
c_n\,=\,{\rme{0}{ p }{n,1}\rme{n,1}{x  }{\text{o-Ps}} \over -i \psi_o(0)} 
&=&   \frac {16} 3 \left[  {  n (n-1)^{n-2} \over (n+1)^{n+2}} \right]. \nn
\label{8.36}
\end{eqnarray}

The continuum states $R_W(r)$ with energy $W=k^2/2$ 
are obtained from the discrete wavefunctions by the replacement
\begin{eqnarray}
n &\to& -i n^\prime,
\label{8.37}
\end{eqnarray}
where
\begin{eqnarray}
n^\prime &=& {Z \over k}.
\label{8.38}
\end{eqnarray}
The exact relation reads
\begin{eqnarray}
R_{nl}(r) &\to &   \left( ik \right)^{3/2} \sqrt{1-e^{-2 \pi n^\prime }} \sqrt{1 \over   Z} R_W(r),  \nn
\label{8.39}
\end{eqnarray}
with $Z=1$ for $e^+e^-$ states.

The completeness relation can be written as
\begin{eqnarray}
&&\sum_n \ket{n} \bra{n} + \int {\rm d}W \ket{W} \bra{W}\nn
&=&\sum_n \ket{n} \bra{n} + \int {\rm d}n^\prime { \ket{n^\prime} \bra{n^\prime} \over  \left(1-e^{-2\pi n^\prime}\right)} .
\label{8.40}
\end{eqnarray}
The sum over intermediate Coulomb states is given by the measure Eq.~(\ref{8.40}), where the continuum states are normalized using the analytic continuation of the discrete states, Eqs.~(\ref{8.37},\ref{8.39}). Note that  the integral over $n^\prime$  in Eq.~(\ref{8.40}) has a non-trivial weight.

The dipole reduced matrix elements for the continuum states read
\begin{eqnarray}
c_{n^{\prime}}&=&{\rme{0}{ p }{n^\prime,1}\rme{n^\prime,1}{x  }{\text{o-Ps}} \over -i \psi_o(0)} \nn
&=&   \frac {16} 3 {  n^\prime \over (1+n^{\prime\, 2})^2 } \exp\left[- \left(\pi -2 \tan^{-1} n^\prime \right) n^\prime \right] .\nn
\label{8.41}
\end{eqnarray}
The $p$ matrix element is given by analytic continuation of Eq.~(\ref{8.35}), and the $x$ matrix element by analytic continuation of Eq.~(\ref{8.31}) plus complex conjugation, since it involves $\bra{n^\prime}$ rather than $\ket{n^\prime}$.

The electric dipole amplitude is (putting back the units):
\begin{eqnarray}
a_e(\omega) 
&=&\sum_{n=2}^\infty {\omega\, c_n  \over \omega+ E_{no} } +
\int_0^\infty { {\rm d} n^\prime  \over 1 - e^{-2 \pi n^\prime} } {\omega\, c_{n^\prime}  \over \omega+ E_{n^\prime o} }\nn
\label{8.42}
\end{eqnarray}
with the energy differences from the ground state 
\begin{eqnarray}
E_{no} &=& \frac 1 4 m\alpha^2\left(1 - {1 \over n^2} \right),  \nn
E_{n^\prime o} &=& \frac 1 4 m\alpha^2 \left(1 + {1 \over n^{\prime\, 2}} \right).
\label{8.43}
\end{eqnarray}
We have verified that Eq.~(\ref{8.42}) is numerically equal to Eq.~(\ref{7.13}) using Eq.~(\ref{7.14}) for the Coulomb Green's function.

\subsection{Decay rate using local annihilation operators} \label{sec:annresult}

\begin{figure}
\begin{center}
\includegraphics[width=5cm]{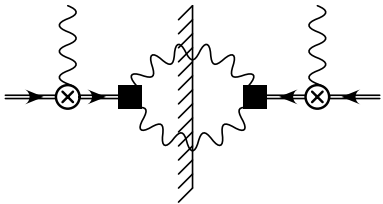}
\caption{Decay rate using four-fermion annihilation operators. The cut graph represents the local annihilation operators.
\label{fig:ann}}
\end{center}
\end{figure}

The decay spectrum Eq.~(\ref{6.18}) can also be computed using the annihilation Lagrangian Eq.~(\ref{annlag}) following the same steps as before. The magnetic dipole transition proceeds via S-wave annihilation, and the electric dipole transition via P-wave annihilation, and the diagrams have the form show in Fig.~\ref{fig:ann}.
 The decay spectrum is
\begin{eqnarray}
{ {\rm d}\Gamma \over  {\rm d}x }  &=&{m \alpha^4 \over 9 \pi^2} x \Biggl[ \text{Im} f_{\gamma\gamma}({}^1S_0)  \abs{a_m}^2 +  \frac 1 3 \text{Im} f_{\gamma\gamma}({}^3 P_0)  \abs{a_e}^2 \nn
&&+  \frac 5 3 \text{Im} f_{\gamma\gamma}({}^3 P_2)  \abs{a_e}^2\Biggr] .
\label{6.18a}
\end{eqnarray}
Using Eq.~(\ref{anncoeff}) reduces this to Eq.~(\ref{6.18}). Of the $7\abs{a_e}^2/3$ electric dipole contribution, the ${}^3 P_0$ contribution is $\abs{a_e}^2$ and the
${}^3 P_2$ contribution is $4 \abs{a_e}^2/3$.

In computing Eq.~(\ref{6.18a}), we have used the matrix element
\begin{eqnarray}
\me{\text{o-Ps}}{ \chi^\dagger_{-\mathbf{p}} \sigma^i \phi_{\mathbf{p}} \ket{0} \bra{0} \phi^\dagger_{\mathbf{p}} \sigma^j \chi_{-\mathbf{p}} }{\text{o-Ps}} &=& 2 \abs{\psi_o(0)}^2 \varepsilon_o^i \varepsilon_o^j \nn
\end{eqnarray}

\section{Results and Conclusions}\label{sec:spectrum}

\begin{figure}
\begin{center}
\includegraphics[width=8cm]{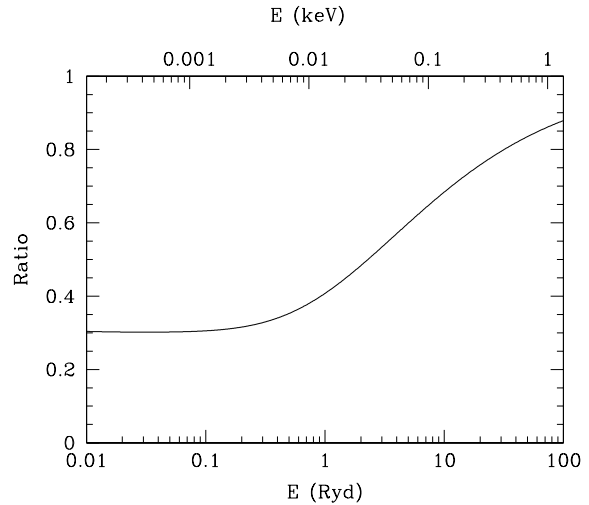}
\caption{Ratio of the o-Ps decay spectrum including binding energy corrections to the Ore-Powell spectrum. The electric dipole amplitude decouples for energies below a Rydberg, leaving only the magnetic contribution of $3/10$.
\label{fig:ratio}}
\end{center}
\end{figure}

\begin{figure}
\begin{center}
\includegraphics[width=8cm]{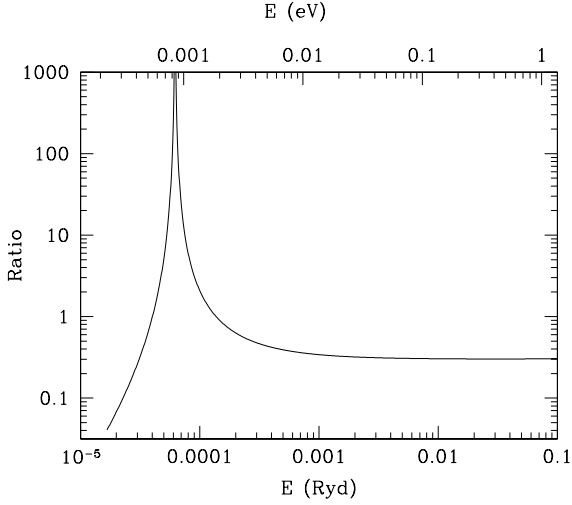}
\caption{Ratio of the o-Ps decay spectrum including binding energy corrections to the Ore-Powell spectrum. The peak is the p-Ps resonance. The amplitude in this region is dominated by the magnetic dipole. For energies below the p-Ps resonance, the ratio to the Ore-Powell spectrum vanishes as $E^2$.
 \label{fig:peak}}
\end{center}
\end{figure}

The o-Ps decay spectrum is given by Eq.~(\ref{6.18}) with $\abs{a_m}^2$ given by Eq.~(\ref{7.03}) and $a_e$ given by Eq.~(\ref{7.13}). The Ore-Powell spectrum is given by Eq.~(\ref{6.18}) with $a_{m,e}$ given by Eq.~(\ref{6.11}). The ratio of the photon spectrum to the Ore-Powell value is shown in Fig.~\ref{fig:ratio}.
At energies large compared with the binding energy, $a_e$ and $a_m$ approach their values in Eq.~(\ref{6.11}), and the spectrum approaches the Ore-Powell spectrum. The magnetic and electric dipole terms contribute in the ratio $3:7$. Note that the approach to the asymptotic value of unity is rather slow. At a photon energy of 100~Ryd, the ratio is $0.88$. At energies small compared with the binding energy, the electric dipole transitions decouple, and one is left with the magnetic contribution of $3/10$ of the Ore-Powell value. The magnetic contribution is given by the p-Ps resonance contribution. The photon decay spectrum in the p-Ps resonance region is shown in Fig.~\ref{fig:peak}.
At energies much smaller than the hyperfine splitting, the p-Ps state also decouples, and the decay rate vanishes as $E_\gamma^3$.

The results are consistent with the Ore-Powell spectrum and with Low's theorem. They include binding effects in a systematic expansion in powers of $v$. The modification of the decay spectrum also leads to a modification of the total decay rate. Since the deviation of the spectrum is of order unity in a region of size the binding energy, and the decay spectrum is of order $x$, the total rate will be corrected at order $\alpha^4$ by the effects we have computed.

\acknowledgments

One of us (AM) would like to thank S.~Freedman and M.~Skalsey for helpful e-discussions. Some of this work was done at the Aspen Center for Physics. AM was supported in part by the Department of Energy under contract  DOE-FG03-97ER40546.
PRF acknowledges support from TMR EURIDICE, EC Contract No.~HPRN-CT-2002-00311, from MCYT (Spain) under grant FPA2001-3031 and from ERDF funds (European Commission).  PRF wishes to thank the high energy physics group at UCSD for his hospitality during his visit there, where part of this work was done.

\begin{appendix}

\section{Parapositronium Decay Amplitude}\label{sec:para}

\begin{figure}
\begin{center}
\includegraphics[width=4cm]{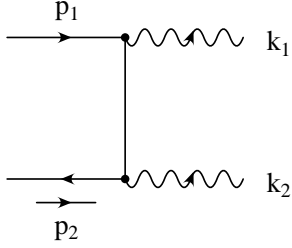}
\end{center}
\caption{Two photon annihilation graph. \label{fig:2photon}}
\end{figure}

The p-Ps$\to 2\gamma$ decay amplitude is given by Fig.~\ref{fig:2photon}.
We require the decay amplitude to first order in $\mathbf{p}$.  The incoming momenta are $p_1=(E,\mathbf{p})$, and  $p_2=(E,-\mathbf{p})$ with $E = \sqrt{m^2 + \mathbf{p}^2} \sim m + \mathcal{O}(p^2)$. The amplitude is
\begin{eqnarray}
A^{(2\gamma)}&=& -ie^2 \bar v \left( p_2 \right) \xslash{\epsilon}_2  {1 \over \xslash{p}_1 - \xslash{k}_1 - m} \xslash{\epsilon}_1  u \left( p_1 \right).\nn
\label{A3}
\end{eqnarray}
We have not included any minus sign coming from Wick contractions 
for the initial $e^+e^-$ state. 
Including the crossed-graph, assuming the polarizations are transverse, and expanding in $\mathbf{p}$ gives
\begin{eqnarray}
A^{(2\gamma)}
&\equiv& \chi^\dagger \left(W_0 + \mathbf{W}_1 \cdot \mathbf{p}\right)\phi
+ {\cal O}(p^2),
\label{A7}
\end{eqnarray}
where
\begin{eqnarray}
W_0&=&
{ e^2 \over 2 m} \left[ \bm{\delta}_1 \cdot \bm{\epsilon}_2 + \bm{\delta}_2 \cdot \bm{\epsilon}_1  \right] ,
\label{A10}
\end{eqnarray}
\begin{eqnarray}
\mathbf{W}_1 &=& { ie^2 \over m^2} \bigg[\left(\bm{\sigma} \cdot \hat{\bm{k}}_1 \right) \left(\bm{\epsilon}_1 \cdot \bm{\epsilon}_2 \right)  \hat{\bm{k}}_1\nn
&& +  \left(\bm{\sigma} \cdot \bm{\epsilon}_2\right)\bm{\epsilon}_1+ 
\left(\bm{\sigma} \cdot \bm{\epsilon}_1 \right)\bm{\epsilon}_2\bigg],
\label{A11}
\end{eqnarray}
and we have used $\bm{k}_1 = - \bm{k}_2$.

\section{Gauge Invariance}\label{sec:gauge}

One can compute in the effective theory using either $\mathbf{x \cdot E}$ or $\mathbf{p \cdot A}$ interactions. We have done the computation using $\mathbf{x \cdot E}$. If one uses $\mathbf{p \cdot A}$ interactions, then the gauge invariant momentum operator with our convention for the covariant derivative is $\mathbf{p}-e\mathbf{A}$. The p-Ps decay amplitude Eq.~(\ref{A7}) is then written in gauge invariant form as
\begin{eqnarray}
W_0 + \mathbf{W}_1 \cdot \left( \mathbf{p}-e\mathbf{A} \right),
\label{B.01}
\end{eqnarray}
and the interaction Hamiltonian Eq.~(\ref{5.01}) has the term
\begin{eqnarray}
H_{\text{int}} &=& - {2 e \over m} \mathbf{p}\cdot \mathbf{A} ,
\label{B.02}
\end{eqnarray}
instead of $-e \mathbf{x} \cdot \mathbf{E}$.
The electric dipole amplitude contribution from Eq.~(\ref{B.02}) is
\begin{eqnarray}
&&{2 e i \over m} \sum_n { \me{0}{ A^{(2\gamma)}}{n} i \me{n}{ \mathbf{p} \cdot \bm{\epsilon}_3 }{\text{o-Ps}}
\over E_o-E_n-E_3}.
\label{B.03}
\end{eqnarray}
Now
\begin{eqnarray}
\left[H,\mathbf{x} \right] &=& - {2i  \over m} \mathbf{p},
\label{B.04}
\end{eqnarray}
so that the amplitude is
\begin{eqnarray}
&& i e  \sum_n  \me{0}{ A^{(2\gamma)}}{n}  \me{n}{ \mathbf{x} \cdot \bm{\epsilon}_3}{\text{o-Ps}}\nn
&& +  i e  \sum_n {E_3 \me{0}{ A^{(2\gamma)}}{n}  \me{n}{ \mathbf{x} \cdot \bm{\epsilon}_3}{\text{o-Ps}}
\over E_o-E_n-E_3} .
\label{B.07}
\end{eqnarray}
The second term agrees with the dipole emission amplitude Eq.~(\ref{6.02}) using the $\mathbf{x \cdot E}$ interaction. The first term gives the extra contribution
\begin{eqnarray}
i e   \me{0}{ A^{(2\gamma)}   \mathbf{x} \cdot \bm{\epsilon}_3}{\text{o-Ps}}
&= & e   \me{0}{ \mathbf{W}_1 \cdot \bm{\epsilon}_3}{\text{o-Ps}}.
\label{B.05}
\end{eqnarray}
In addition, one has the amplitude from the $-e\mathbf{A}$ term in Eq.~(\ref{B.01}),
\begin{eqnarray}
-e \me{0}{ \mathbf{W}_1 \cdot \bm{\epsilon}_3}{\text{o-Ps}}
\label{B.06}
\end{eqnarray}
which exactly cancels Eq.~(\ref{B.05}), so the net contribution is gauge invariant.

\section{Coulomb Green's Function}\label{sec:coulomb}

The Coulomb Green's function satisfies the differential equation
\begin{eqnarray}
\left( H_o + {k^2 \over m} \right)   G\left( \mathbf{x}, \mathbf{y}, k \right) = \delta\left( \mathbf{x} - \mathbf{y} \right),
\label{C1}
\end{eqnarray}
with the Coulomb Hamiltonian for positronium
\begin{eqnarray}
H_o = {\mathbf{p}^2 \over m} - {\alpha \over r} .
\label{C2}
\end{eqnarray}
In terms of a complete set of states in position space,
\begin{eqnarray}
G\left( \mathbf{x}, \mathbf{y}, k \right)  &=& \sum_n { \psi_n \left( \mathbf{x} \right)  \psi_n^* \left( \mathbf{y} \right) \over E_n + k^2/m}  .
\label{C3}
\end{eqnarray}
The Green's function admits a partial wave decomposition:
\begin{eqnarray}
G\left( \mathbf{x}, \mathbf{y}, k \right)  &=& \sum_{\ell=0}^\infty \left(2 \ell +1 \right) \left( x y \right)^\ell
P_\ell \left( \mathbf{x \cdot y} / x y \right) G_\ell \left( x, y ,k \right)\nn
\label{C4}
\end{eqnarray}
where $P_\ell(x)$ are the Legendre polynomials.  The partial waves of the Green's function are~\cite{hostler}
\begin{eqnarray}
&&G_\ell \left( x, y ,k \right) \nn
&=& {m k \over 2 \pi} \left(2 k \right)^{2\ell} e^{-k( x + y) } \sum_{r=0}^\infty{ L_r^{2\ell+1}(2 k x)  L_r^{2\ell+1}(2 k y) r! \over (r+\ell+1-\nu)(r+2\ell+1)!} \nn
\label{C5}
\end{eqnarray}
with the parameter
\begin{eqnarray}
\nu = { \alpha m \over 2 k},
\label{C6}
\end{eqnarray}
and the associated Laguerre polynomials defined by
\begin{eqnarray}
L_r^k(x) &=& {e^x x^{-k} \over r!} \left( { {\rm d} \over {\rm d} x} \right)^r e^{-x} x^{r+k}.
\label{C7}
\end{eqnarray}

The Green's function partial waves have a more compact expression when one of the arguments is zero~\cite{penin}. Set $y=0$ and use
\begin{eqnarray}
L_r^k(0) &=& {(r+k)! \over r!\, k!},
\label{C8}
\end{eqnarray}
to get
\begin{eqnarray}
&&G_\ell \left( x, 0 ,k \right) \nn
&=& {m k \over 2 \pi} \left(2 k \right)^{2\ell} e^{-k x } \sum_{r=0}^\infty{ L_r^{2\ell+1}(2 k x)   \over (r+\ell+1-\nu)(2\ell+1)!} .\nn
\label{C9}
\end{eqnarray}
The generating function for the associated Legendre Polynomials~\cite{arfken} is
\begin{eqnarray}
\sum_{r=0}^\infty L_r^a(u) \, s^r &=& {e^{-u s/(1-s)} \over (1-s)^{a+1}}.
\label{C10}
\end{eqnarray}
Also
\begin{eqnarray}
{1 \over r+b} &=& \int_0^1 {\rm  d} s\ s^{r+b-1}.
\label{C11}
\end{eqnarray}
The identities (\ref{C10}) and (\ref{C11}) allow us to write
\begin{eqnarray}
\sum_{r=0}^\infty {L_r^a(u) \over r+b } &=& \int_0^1 {\rm  d} s\ s^{b-1} {e^{-u s/(1-s)} \over (1-s)^{a+1}}.
\label{C12}
\end{eqnarray}
Changing variables to $t=s/(1-s)$ gives
\begin{eqnarray}
\sum_{r=0}^\infty {L_r^a(u) \over r+b }
&=& \int_0^\infty {\rm  d} t \, t^{b-1} (1+t)^{a-b} e^{-u t}\nn
&=& \Gamma(b) \, U(b,a+1,u),
\label{C15}
\end{eqnarray}
where we have adopted the usual definition of
the confluent hypergeometric function $U(a,b,z)$, see e.g. 
Ref.~\cite{abramowitz}. 
Applied to the sum in Eq.~(\ref{C9}), the result~(\ref{C15}) yields 
\begin{eqnarray}
&&G_\ell \left( x, 0 ,k \right) \nn
&=& {m k \over 2 \pi} \left(2 k \right)^{2 \ell} e^{-k x }
{\Gamma(\ell +1-\nu)U(\ell+1-\nu,2\ell+2,2kx)\over (2\ell+1)!} \nn
\label{C16}
\end{eqnarray}
which differs from Ref.~\cite{penin} by the $(2\ell+1)!$ factor in the
denominator.

\end{appendix}


\begin{thebibliography}{1}

\bibitem{deutsch} M.~Deutsch, Phys.~Rev.~82 (1951) 455.

\bibitem{pestieau} J.~Pestieau and C.~Smith, Phys.~Lett.~B524 (2002) 395.

\bibitem{OrePowell} A.~Ore and J.L.~Powell, Phys.~Rev.~75 (1949) 1696.

\bibitem{caswell}
W.~E.~Caswell and G.~P.~Lepage,
Phys.\ Lett.\ B {\bf 167}, 437 (1986).

\bibitem{bbl}
G.~T.~Bodwin, E.~Braaten and G.~P.~Lepage,
Phys.\ Rev.\ D {\bf 51}, 1125 (1995)
[Erratum-ibid.\ D {\bf 55}, 5853 (1997)].

\bibitem{lmr}
M.~E.~Luke, A.~V.~Manohar and I.~Z.~Rothstein,
Phys.\ Rev.\ D {\bf 61}, 074025 (2000).

\bibitem{ps}
A.~Pineda and J.~Soto,
Nucl.\ Phys.\ Proc.\ Suppl.\  {\bf 64}, 428 (1998).


\bibitem{lms}
M.~E.~Luke and A.~V.~Manohar,
Phys.\ Rev.\ D {\bf 55}, 4129 (1997);
A.~V.~Manohar and I.~W.~Stewart,
Phys.\ Rev.\ Lett.\  {\bf 85}, 2248 (2000).

\bibitem{low}
F.~E.~Low,
Phys.\ Rev.\  {\bf 110}, 974 (1958).

\bibitem{itzykson} C.~Itzykson and J.~Zuber, {\sl Quantum Field Theory},
McGraw-Hill (1980). 

\bibitem{cohen} C.~Cohen-Tannoudji, J.~Dupont-Roc and G.~Grynberg,
{\it Photons \& Atoms, Introduction to Quantum Electrodynamics}, 
John Wiley \& Sons (1989). 

\bibitem{pinedasoto}
A.~Pineda and J.~Soto,
Phys.\ Lett.\ B {\bf 420}, 391 (1998).

\bibitem{multipole}
P.~Labelle,
Phys.\ Rev.\ D {\bf 58}, 093013 (1998);
B.~Grinstein and I.~Z.~Rothstein,
Phys.\ Rev.\ D {\bf 57}, 78 (1998).

\bibitem{yang}
C.~N.~Yang,
Phys.\ Rev.\  {\bf 77} 242 (1950).



\bibitem{wigner} E.P.~Wigner, {\sl Gruppentheorie und ihre Anwendung
auf die Quantenmechanik der Atomspektren}, Friedrich Vieweg und Sohn (1931);
C.~Eckart, Rev.~Mod.~Phys.~2 (1930) 305.

\bibitem{penin} A.~A.~Penin and A.~A.~Pivovarov,
Nucl.~Phys.~B550 (1999) 375.

\bibitem{voloshin}
M.~B.~Voloshin,
arXiv:hep-ph/0311204.

\bibitem{bethe} H.A.~Bethe and E.E.~Salpeter, {\sl Quantum Mechanics of One- and Two-Electron Atoms}, Academic Press (1957).

\bibitem{hostler} L.C.~Hostler, J.~Math.~Phys.~11 (1970) 2966.

\bibitem{arfken} G.~Arfken, {\sl Mathematical Methods for Physicists}, 
Academic Press (1970).

\bibitem{abramowitz} M.~Abramowitz and I.A.~Stegun, {\sl Handbook of
Mathematical Functions}, Dover Publications (1970).

 
\end{thebibliography}
\end{document}